\documentstyle[pra,aps]{revtex}

\begin{document}
\title{Low energy collective excitations in a superfluid trapped Fermi gas}
\author{M. A. Baranov and D. S. Petrov}
\date{\today }
\address{Russian Research Center Kurchatov Institute,\\
Kurchatov Square, 123182 Moscow, Russia}
\maketitle

\begin{abstract}
We study low energy collective excitations in a trapped superfluid Fermi
gas, that describe slow variations of the phase of the superfluid order
parameter. Well below the critical temperature the corresponding
eigenfrequencies turn out to be of order the trap frequency, and these modes
manifest themselves as the eigenmodes of the density fluctuations of the gas
sample. The latter could provide an experimental evidence of the presence of
the superfluid phase.
\end{abstract}

PACS number(s): 03.75.Fi, 05.30.Fk

\vspace{4mm}

The search of a novel behavior of a gaseous system in the confined geometry
of a trap attracts a lot of attention since the discovery of BEC in trapped
gas samples of alkali atoms \cite{Cor95,Hul95,Ket95}. This behavior is
mostly related to the combined effects of an interparticle interaction and
quantum statistics, which plays a dominated role below the degeneracy
temperature. In the case of a Bose system, the latter corresponds to the
critical temperature of BEC. For a Fermi system the degeneracy temperature
equals to the Fermi energy $T_{F}=\varepsilon _{F}$, and at temperatures
below $T_{F}$ the Pauli exclusion principle should be taken into account 
\cite{Zol,Vic,DeM}. Remarkably, that at much lower temperature (called the
critical temperature $T_{c}\ll T_{F}$) there could be a transition into a
superfluid phase, as a result of Cooper pairing. Possible versions of the
transition have recently been discussed in Refs.\cite{HS,BKK,S,BP}. Although
the pairing strongly influences only a small fraction $(\sim
T_{c}/\varepsilon _{F}\ll 1)$ of quantum states in the vicinity of the
chemical potential, it nevertheless changes the behavior of the Fermi system
significantly, because precisely those states govern the response of the
system to a small external perturbation. For example, as it was shown in
Ref. \cite{B}, the superfluid pairing changes (smears out) the resonance
structure of the gas density oscillations in a parabolic trap.

In this paper we study low energy collective excitations in a superfluid
Fermi gas trapped in an isotropic harmonic potential. These excitations
correspond to the phase fluctuations of the order parameter $\Delta ({\bf r}%
) $ and, as shown, well-below the transition temperature their
eigenfrequencies are of the order of the trap frequency. The damping of the
excitations is expected to be small, and, therefore, these modes are
well-defined collective excitations. They manifest themselves as
oscillations of the superfluid component of the Fermi gas, and hence, at low
enough temperature, when the density of a normal component tends to zero
exponentially, these collective excitations define the eigenmodes for the
density fluctuations of the entire gas sample. As these modes exist only
below the transition temperature, their experimental observation could
provide an indication of the presence of the superfluid phase.

Specific properties of the collective modes in a superfluid phase depend on
the structure of the order parameter and, hence, on the type of pairing. For
a singlet $s$-wave pairing the order parameter is a complex function, and
one has two branches of collective excitations corresponding to the phase
and modulus variations of the order parameter. For a triplet $p$-wave
pairing the order parameter is a complex $3\times 3$ matrix, and, hence,
additional branches of collective modes appear (see, e.g., Ref.\cite{Wolfle}%
). However, in both cases the lowest in energy branch corresponds to the
fluctuations of the phase of the order parameter (Bogolyubov sound). In this
paper we study this mode for a trapped superfluid Fermi gas. For simplicity
we consider the case with a ''singlet'' $s$-wave pairing, that implies the
presence of an attractive interatomic interaction. This situation can be
realized in a gas of $^{6}$Li atoms in a magnetic trap, where they are
characterized by a large and negative triplet $s$-wave scattering amplitude $%
a\approx -1140\AA $ \cite{H} between the two hyperfine components.

We consider a two component gas of fermionic atoms ($\alpha $ and $\beta $
atoms) trapped in an isotropic harmonic potential. We assume the two
components (for example, hyperfine components) have equal masses and
concentrations. The Hamiltonian of the system is ($\hbar =1$) 
\begin{equation}
H=\sum_{i=\alpha ,\beta }\int d{\bf r}\psi _{i}^{\dagger }({\bf r})H_{0}\psi
_{i}({\bf r})+V\int d{\bf r}\psi _{\alpha }^{\dagger }({\bf r})\psi _{\alpha
}({\bf r})\psi _{\beta }^{\dagger }({\bf r})\psi _{\beta }({\bf r}),
\label{1}
\end{equation}
where $\psi _{i}({\bf r})$ with $i=\alpha ,\beta $ are the field operators
of $\alpha $ and $\beta $ atoms, $H_{0}=-\nabla ^{2}/2m+m\Omega
^{2}r^{2}/2-\mu $, $\Omega $ the trap frequency, $m$ the mass of the atoms,
and $\mu $ the chemical potential. The second term in Eq. (\ref{1})
corresponds to an attractive elastic short-range interaction between $\alpha 
$ and $\beta $ atoms ($s$-wave scattering length $a<0$) with $V=4\pi a/m$
being the coupling constant. In the Hamiltonian (\ref{1}) we neglect the
interaction between atoms of the same sort, originating in the case of
fermions only from the scattering with orbital angular momentum $l\geq 1$.
The presence of an attractive intercomponent interaction in the $s$-wave
scattering channel leads to a superfluid phase transition \cite{S} with the
critical temperature $T_{c}\ll \mu $. We assume that $T_{c}$ is much larger
than $\Omega $ and, hence, the value of the critical temperature in the trap
is very close \cite{BP} to the critical temperature $T_{c}^{(0)}$ in a
spatially homogeneous gas with the density being equal to the maximum
density $n_{0}$ of the trapped gas sample, $T_{c}^{(0)}=0.28\,\varepsilon
_{F}\exp \{-1/\lambda \}$ \cite{GM-B}, where $\lambda =2|a|p_{F}/\pi $ is
the small parameter of the theory ($\lambda <1$), $p_{F}=(3\pi
^{2}n_{0})^{1/3}$ and $\varepsilon _{F}=p_{F}^{2}/2m$.

The superfluid phase is characterized by the order parameter (complex
function) $\Delta ({\bf r})=\left| V\right| \,\left\langle \psi _{\alpha }(%
{\bf r})\psi _{\beta }({\bf r})\right\rangle $. The equilibrium form of $%
\Delta ({\bf r})$ for a trapped gas sample was studied analytically in Ref. 
\cite{B} and numerically in Ref. \cite{Bru1}. The appearance of $\Delta (%
{\bf r})$ strongly influences only the quantum states within the range of
order $T_{c}$ near the chemical potential level. As a result, the gas
density profile changes only slightly ($\Delta n({\bf r})/n({\bf r})\sim
T_{c}/\mu \ll 1$), and hence, one has $n({\bf r}%
)=n_{0}(1-(r/R_{TF})^{2})^{3/2}$ in the Thomas-Fermi approximation for both
normal and superfluid phases ($R_{TF}=p_{F}/m\Omega $ is the Thomas-Fermi
radius of the gas cloud). The interparticle interaction leads to corrections
to this formula. However, the leading the mean-field corrections \cite{Bru}
are proportional to the small parameter of the theory $\lambda $, and hence,
will be neglected.

At finite temperatures it is convenient to study the evolution of the system
in imaginary (Matsubara) time $\tau \in \lbrack 0,1/T]$ \cite{LL}. If one
splits the order parameter $\Delta ({\bf r},\tau )$ into its equilibrium
part $\Delta _{0}({\bf r})$ ($\Delta _{0}^{\ast }({\bf r})=\Delta _{0}({\bf r%
})$) and a small fluctuation $\delta ({\bf r},\tau )=T\sum_{\omega }\delta
_{\omega }({\bf r})\exp (-i\omega \tau )$, where $\omega =\pi T(2n+1)$ with $%
n$ being an integer, is the Matsubara frequency, then the equation for $%
\delta _{\omega }({\bf r})$ reads (see, e.g. \cite{Popov} ) 
\begin{equation}
\frac{\delta _{\omega }({\bf r})}{\left| V\right| }=T\sum_{\omega _{1}}\int d%
{\bf r}^{\prime }\{G_{\omega +\omega _{1}}({\bf r,r}^{\prime })G_{-\omega
_{1}}({\bf r,r}^{\prime })\delta _{\omega }({\bf r}^{\prime })-F_{\omega
+\omega _{1}}({\bf r,r}^{\prime })F_{\omega _{1}}({\bf r,r}^{\prime })\delta
_{\omega }^{\ast }({\bf r}^{\prime })\}  \label{2}
\end{equation}
(the field $\delta _{\omega }^{\ast }({\bf r})$ obeys a complex conjugate
equation). In Eq. (\ref{2}) $G_{\omega }({\bf r,r}^{\prime })$, $F_{\omega }(%
{\bf r,r}^{\prime })$ are the normal and anomalous Green functions of the
Hamiltonian 
\[
\widetilde{H}_{0}=\int d{\bf r}\left\{ \sum_{i=\alpha ,\beta }\psi
_{i}^{\dagger }H_{0}\psi _{i}+\Delta _{0}({\bf r})(\psi _{\alpha }\psi
_{\beta }+\psi _{\beta }^{\dagger }\psi _{\alpha }^{\dagger })\right\} 
\]
in the Matsubara representation. They can be expressed through solutions ($%
u_{\nu }({\bf r}),v_{\nu }({\bf r})$) of the Bogolyubov-de Gennes equations 
\begin{equation}
H_{0}\left( 
\begin{array}{c}
u_{\nu }({\bf r}) \\ 
v_{\nu }({\bf r})
\end{array}
\right) +\Delta _{0}({\bf r})\left( 
\begin{array}{c}
v_{\nu }({\bf r}) \\ 
-u_{\nu }({\bf r})
\end{array}
\right) =\varepsilon _{\nu }\left( 
\begin{array}{c}
u_{\nu }({\bf r}) \\ 
-v_{\nu }({\bf r})
\end{array}
\right) ,  \label{3}
\end{equation}
which define the Bogolyubov (unitary) transformation from the bare fermions $%
\psi _{\alpha }$, $\psi _{\beta }$ to single-particle excitations $\alpha
_{\nu }$, $\beta _{\nu }$with energies $\varepsilon _{\nu }\geq 0$

\begin{equation}
\left( 
\begin{array}{c}
\psi _{\alpha }({\bf r}) \\ 
\psi _{\beta }({\bf r})
\end{array}
\right) =\sum_{\nu }\left[ u_{\nu }({\bf r})\left( 
\begin{array}{c}
\alpha _{\nu } \\ 
\beta _{\nu }
\end{array}
\right) +v_{\nu }^{\ast }({\bf r})\left( 
\begin{array}{c}
\beta _{\nu }^{\dagger } \\ 
-\alpha _{\nu }^{\dagger }
\end{array}
\right) \right] ,  \label{4}
\end{equation}
such that the Hamiltonian $\widetilde{H}_{0}$ is diagonal. Then the Green
functions are 
\[
G_{\omega }({\bf r}_{1},{\bf r}_{2})=\sum_{\nu }\left\{ \frac{u_{\nu }({\bf r%
}_{1})u_{\nu }^{\ast }({\bf r}_{2})}{i\omega -\varepsilon _{\nu }}+\frac{%
v_{\nu }^{\ast }({\bf r}_{1})v_{\nu }({\bf r}_{2})}{i\omega +\varepsilon
_{\nu }}\right\} , 
\]
\begin{equation}
F_{\omega }({\bf r}_{1},{\bf r}_{2})=\sum_{\nu }\left\{ \frac{u_{\nu }({\bf r%
}_{1})v_{\nu }^{\ast }({\bf r}_{2})}{i\omega -\varepsilon _{\nu }}-\frac{%
v_{\nu }^{\ast }({\bf r}_{1})u_{\nu }({\bf r}_{2})}{i\omega +\varepsilon
_{\nu }}\right\} .  \label{5}
\end{equation}

For the fluctuation of the phase of the order parameter one has $\delta
_{\omega }({\bf r})=-\delta _{\omega }^{\ast }({\bf r})=2i\Delta _{0}({\bf r}%
)\varphi _{\omega }({\bf r})$ with $\varphi _{\omega }({\bf r})\ll 1$ being
a real function, and Eq. (\ref{2}) can be rewritten in the form 
\begin{equation}
\frac{\delta _{\omega }(r)}{\left| V\right| }-\int d{\bf r}^{\prime }K_{0}(%
{\bf r,r}^{\prime })\delta _{\omega }({\bf r}^{\prime })=\int d{\bf r}%
^{\prime }K_{\omega }({\bf r,r}^{\prime })\delta _{\omega }({\bf r}^{\prime
}),  \label{6}
\end{equation}
where $K_{0}=T\sum_{\omega _{1}}(G_{\omega _{1}}G_{-\omega _{1}}+F_{\omega
_{1}}F_{\omega _{1}})$ is the static kernel and $K_{\omega }=T\sum_{\omega
_{1}}(G_{\omega +\omega _{1}}G_{-\omega _{1}}+F_{\omega +\omega
_{1}}F_{\omega _{1}})-K_{0}$ the dynamic one.

For $(T_{c}-T)/T_{c}\ll 1$ the order parameter is small, $\Delta ({\bf r}%
)\ll T_{c}$, and one can neglect the contribution from the anomalous Green
functions. Under the assumption $T_{c}\gg \Omega $, the normal phase Green
functions $G_{\omega }({\bf r,r}^{\prime })$ can be taken in the
semiclassical approximation. Then for a slow varying function $\varphi $ one
gets the following results for the left and right hand sides of Eq. (\ref{6}%
) 
\begin{equation}
\frac{\delta _{\omega }({\bf r})}{\left| V\right| }{\bf r}-\int d{\bf r}%
^{\prime }K_{0}({\bf r,r}^{\prime })\delta _{\omega }({\bf r}^{\prime
})=2iN(R)\frac{7\zeta (3)}{96\pi ^{2}}\frac{\Omega ^{2}}{T_{c}^{2}}\left( 
\frac{\Delta _{0}}{\sqrt{1-R^{2}}}\nabla _{{\bf R}}\left[
(1-R^{2})^{3/2}\nabla _{{\bf R}}\varphi \right] +2(1-R^{2})\nabla _{{\bf R}%
}\Delta _{0}\nabla _{{\bf R}}\varphi \right) ,  \label{7}
\end{equation}
where $\zeta (z)$ is the Riemann zeta function, $N(R)=(mp_{F}/\pi ^{2})\sqrt{%
1-R^{2}}$ the density of states on a local Fermi surface ($R=r/R_{TF}$), and

\begin{equation}
\int d{\bf r}^{\prime }K_{\omega }({\bf r,r}^{\prime })\delta _{\omega }(%
{\bf r}^{\prime })=2i\Delta _{0}{\bf (}R)N(R)\frac{\pi \left| \omega \right| 
}{16T_{c}}\varphi .  \label{8}
\end{equation}

For $T\ll T_{c}$ the calculations are more laborious. In this case one can
use, for example, Eqs. (\ref{5}) and various relations between $u_{\nu }$
and $v_{\nu }$, that follows from Eqs.. (\ref{3}) and from unitarity of the
Bogolyubov transformation, Eq. (\ref{4}). The answer for the left hand side
of Eq. (\ref{6}) is 
\begin{equation}
\frac{\delta _{\omega }({\bf r})}{\left| V\right| }-\int d{\bf r}^{\prime
}K_{0}({\bf r,r}^{\prime })\delta _{\omega }({\bf r}^{\prime })=2i\frac{%
\Omega ^{2}}{4mv_{F}^{2}\Delta _{0}(R)}\nabla _{{\bf R}}\left[ n(R)\nabla _{%
{\bf R}}\varphi \right] .  \label{9}
\end{equation}

The kernel $K_{\omega }({\bf r,r}^{\prime })$ can be calculated by using
Eqs. (\ref{5}) with the semiclassical solutions of the Bogolyubov-de Gennes
equations (\ref{3}), which were found in Ref. \cite{B} for the considered
case of a spherically symmetric harmonic trapping potential. In this case
there are two different types of excitations: in-gap and above-gap. The
eigenenergies of the in-gap excitations are smaller than the maximum value
of the order parameter $\Delta (0)$, and their wave functions $(u_{\nu }(%
{\bf r}),v_{\nu }({\bf r}))$ are pushed by $\Delta ({\bf r})$ to the outer
part of the gas sample \cite{B}. For this reason these states give only
exponentially small contribution $(\sim \exp (-T_{c}/\Omega ))$ to the right
hand side of Eq. (\ref{6}), and, hence, only above-gap excitations are
important. The calculations with slow varying function $\varphi $ then yield 
\begin{equation}
\int d{\bf r}^{\prime }K_{\omega }({\bf r,r}^{\prime })\delta _{\omega }(%
{\bf r}^{\prime })=-2i\omega ^{2}\frac{N(R)}{4\Delta _{0}(R)}\varphi .
\label{10}
\end{equation}
(This result can also be obtained by using the Green functions $G_{\omega }(%
{\bf r,r}^{\prime })$ and $F_{\omega }({\bf r,r}^{\prime })$ in the local
density approximation.)

After making an analytic continuation in Eqs. (\ref{8}) and (\ref{10}) from
the Matsubara frequency $\omega $ to a real one $E$, and combining the
results with Eqs. (\ref{7}) and (\ref{9}), we finally obtain the equations
describing the collective modes related to the phase fluctuations of the
order parameter (Bogolyubov sound).

For $(T_{c}-T)/T_{c}\ll 1$ one has 
\begin{equation}
-\frac{7\Omega ^{2}\zeta (3)}{6\pi ^{3}T_{c}}\left( \frac{1}{\sqrt{1-R^{2}}}%
\nabla _{{\bf R}}\left[ (1-R^{2})^{3/2}\nabla _{{\bf R}}\varphi ({\bf R}%
)\right] +2(1-R^{2})\nabla _{{\bf R}}\ln \Delta _{0}\nabla _{{\bf R}}\varphi
({\bf R})\right) =iE\varphi ({\bf R}),  \label{11}
\end{equation}
and it follows from this equation that for temperatures close to the
critical temperature $T_{c}$ the eigenenergies $E$ are purely imaginary: the
collective mode decays rapidly into pairs of single-particle excitations.

For $T\ll T_{c}$ the equation reads 
\begin{equation}
-\frac{\Omega ^{2}}{3}\frac{1}{\sqrt{1-R^{2}}}\nabla _{{\bf R}}\left[
(1-R^{2})^{3/2}\nabla _{{\bf R}}\varphi ({\bf R})\right] =E^{2}\varphi ({\bf %
R}),  \label{12}
\end{equation}
and in this case the eigenenergies $E$ of the collective modes are real and
of order the trap frequency $\Omega $. These modes being excited results in
oscillations of the superfluid current ${\bf j}=(n/m)\nabla \varphi $ and
density $n=n_{0}+\delta n$, that are related to each other by the continuity
equation $\partial n/\partial t+{\rm div}{\bf j}=0$. (For $T\ll T_{c}$ one
can neglect the contribution of the normal component of the gas and hence, $%
{\bf j}_{s}{\bf =j}$ and $n_{s}=n$.) As a result, the entire gas sample
oscillates according to the formula

\[
n({\bf r},t)=n_{0}({\bf r})+\delta n({\bf r},t)\approx \left[ 1+\frac{1}{m}%
\nabla ^{2}\psi \right] \ \ n_{0}({\bf r}+\frac{1}{m}\nabla \psi ), 
\]
where $\psi ({\bf r},t)=\int^{t}\varphi ({\bf r},t^{\prime })dt^{\prime }$%
.(The prefactor in the square brackets ensures the conservation of a total
number of particles.)

The damping of the collective modes, which does not contained in Eq. (\ref
{12}), will be mainly determined by processes of decay and scattering on
in-gap excitations (these mechanisms of damping of collective modes in
Bose-condensed atomic gases are often called as Beliaev and
Szepfalusy-Kondor mechanisms respectively; see, e.g. \cite{Gora}). One can
say that the energy of the collective mode converts into a normal component
generated in the outer part of the gas sample. But because the coupling
between the order parameter fluctuations and the in-gap excitations is
exponentially weak $(\sim \exp (-T_{c}/\Omega ))$ , the damping is expected
to be small.

It should be mentioned that Eq. (\ref{12}) can be obtained in the framework
of a hydrodynamic description of a superfluid Fermi gas. When the superfluid
velocity ${\bf v}_{s}=m^{-1}\nabla \varphi $ and the deviation of the
particle density $\delta n$ from its equilibrium value $n_{0}$ are small,
the corresponding Hamiltonian has the form

\begin{equation}
H_{{\rm h}}=\int d{\bf r}\left\{ \frac{1}{2m}n(\nabla \varphi
)^{2}+U(n_{0}+\delta n)\right\} \approx \int d{\bf r}\left\{ \frac{1}{2m}%
n_{0}(\nabla \varphi )^{2}+\frac{1}{2}U^{\prime \prime }(n_{0})\delta
n^{2}+U(n_{0})\right\} ,  \label{13}
\end{equation}
where $U(n)$ is the density dependent part of the energy. (The equilibrium
density $n_{0}$ is defined by the condition $U^{\prime }(n_{0})=0$.) In the
Thomas -Fermi approximation,

\begin{equation}
U(n)=2\int_{p\leq p_{F}(n)}\frac{p^{2}}{2m}\frac{d{\bf p}}{(2\pi )^{3}}%
+\left( \frac{m\Omega ^{2}r^{2}}{2}-\mu \right) n=\frac{3}{10}(3\pi
^{2})^{2/3}\frac{n^{5/3}}{m}+\left( \frac{m\Omega ^{2}r^{2}}{2}-\mu \right)
n,  \label{14}
\end{equation}
and the equilibrium gas density profile $n_{0}(r)=(p_{F}^{3}/3\pi ^{2})\sqrt{%
1-R^{2}}$, as it should be. (In Eq. (\ref{14}) we do not include the effects
of the mean field interaction and the superfluid pairing because they
contain small parameters $\lambda $ and $(T_{c}/\varepsilon _{F})^{2}$
respectively.) For the quantity $U^{\prime \prime }(n_{0})$ in Eq. (\ref{13}%
) one now has $U^{\prime \prime }(n_{0})=(3\pi ^{2})^{-2/3}N(r)^{-1}$, and
the standard commutation relation $\left[ \delta n({\bf r}_{1}),\varphi (%
{\bf r}_{2})\right] =i\delta ({\bf r}_{1}-{\bf r}_{2})$ leads to the
equations

\begin{eqnarray*}
\partial \varphi /\partial t &=&i\left[ H_{{\rm h}},\varphi \right]
=U^{\prime \prime }(n_{0})\delta n \\
\partial (\delta n)/\partial t &=&i\left[ H_{{\rm h}},\delta n\right]
=-\nabla (n_{0}\nabla \varphi )/m,
\end{eqnarray*}
from which one immediately gets Eq. (\ref{12}) for the phase dynamics and
the equation

\begin{equation}
\partial ^{2}(\delta n)/\partial t^{2}+\frac{\Omega ^{2}}{3}\nabla _{{\bf R}%
}\left[ (1-R^{2})^{3/2}\nabla _{{\bf R}}\frac{\delta n}{\sqrt{1-R^{2}}}%
\right] =0  \label{15}
\end{equation}
for the density fluctuations.

Equation (\ref{12}) (or (\ref{15})) together with the condition that $%
\varphi $ (or $\delta n$) is finite, provides us with the eigenfrequencies
and eigenfunctions of the collective modes of the superfluid Fermi gas. For
a spherically symmetric breathing modes (angular momentum $l$ is zero) one
has for the eigenfrequencies

\begin{equation}
(E_{n0}/\Omega )^{2}=\frac{4}{3}n(n+2),\ n=1,2,\ldots  \label{16}
\end{equation}
($n=0$ corresponds to a constant phase and, hence, has no physical
significance), and for the eigenfunctions

\begin{equation}
\varphi _{n0}({\bf R})\propto \ _{2}F{}_{1}(-n,n+2;\frac{3}{2};R^{2}),
\label{17}
\end{equation}
where $_{2}F_{1}(a,b;c;z)$ is the hypergeometric function.

For nonzero angular momentum $l$,

\begin{equation}
(E_{n0}/\Omega )^{2}=l+\frac{4}{3}n(n+l+2),\ n=0,1,2,\ldots  \label{18}
\end{equation}
and

\begin{equation}
\varphi _{nl}({\bf R})\propto R^{l}\ _{2}F{}_{1}(-n,n+l+2;\frac{3}{2}%
+l;R^{2}){\rm Y}_{lm}(\theta ,\phi ).  \label{19}
\end{equation}
(Note, that the eigenfunctions (\ref{17}) and (\ref{19}) are orthogonal with
the weight $1/\sqrt{1-R^{2}}$.)

The spectrum (\ref{16}) and (\ref{18}) coincide with that for a trapped
normal Fermi gas in a hydrodynamic regime \cite{BC}, and the lowest
eigenfrequencies are equal to those for a classical gas \cite{Str1}. The
lowest eigenmodes are of a special interest because they can be excited by a
modulation of the trap frequency. (A small external perturbation $V_{{\rm ext%
}}$ results in an extra term $-iEV_{{\rm ext}}$, or $\partial V_{{\rm ext}%
}/\partial t$ in the time representation, on the right hand side of Eq. (\ref
{12}).) It is reasonable to assume that the trapped Fermi gas just above $%
T_{c}$ is in a collisionless regime ($\Omega \tau \sim \Omega (\lambda
^{2}T_{c}^{2}/\varepsilon _{F})^{-1}\sim (\Omega /\lambda ^{2}T_{c})\exp
(1/\lambda )\gg 1$), where the lowest monopole, dipole and quadrupole
collective modes were calculated in \cite{VicciStr}. For the superfluid
phase, as follows from Eqs. (\ref{16}) and (\ref{18}), the lowest
eigenfrequency $E_{10}$ for the monopole breathing mode ($l=0$, $n=1$) is
equal to $2\Omega $ (this result can be obtained on the basis of the sum
rules \cite{Legg}), and one has the anticipated result $E_{01}=\Omega $ for
a dipole mode $l=1$, $n=0$. These eigenfrequencies coincide (within the
considered accuracy) with those from Ref. \cite{VicciStr}. On the other
hand, Eq. (\ref{18}) gives $E_{02}=\sqrt{2}\Omega $ for the lowest
quadrupole mode, while one has $E_{02}=2\Omega $ \cite{VicciStr} in the
collisionless regime above $T_{c}$. Experimentally the quadrupole mode can
be excited by a small anti-phase modulation of the trap frequency in, for
example, $x$ and $y$ directions, $V_{{\rm ext}}({\bf r},t)=(m\Omega
^{2}/2)(x^{2}-y^{2})\zeta \cos (\omega t)$ with $\zeta \ll 1$, and the
response of the gas sample for $T>T_{c}$ and $T\ll T_{c}$ will have
resonances at different frequencies, $2\Omega $ and $\sqrt{2}\Omega $
respectively.

In conclusion, we have found the low energy collective modes in the
superfluid trapped Fermi gas. These modes are related to the fluctuations of
the phase of the superfluid order parameter, and, hence, describe the motion
of the superfluid component of the gas. Just below the critical temperature
of the superfluid phase transition the eigenenergies of the collective modes
are purely imaginary, and these modes describe the diffusive relaxation of a
superfluid fluctuation. For temperatures well below $T_{c}$, the
eigenenergies are of order the trap frequency, and the damping is small.
Therefore, these modes can manifest themselves as the eigenmodes of the gas
density oscillations that can be observed experimentally, and serve as an
indication of the superfluid phase transition.

We acknowledge fruitful discussions with A.J. Leggett, G.V. Shlyapnikov and
L. Vichi. This work was supported by the Stichting voor Fundamenteel
Onderzoek der Materie (FOM), by INTAS (grant 97.0972), and by the Russian
Foundation for Basic Studies (grant 97-02-16532).

\end{document}